\documentclass[aps,prb,superscriptaddress,showpacs,citeautoscript,twocolumn]{revtex4}

\usepackage{graphicx}
\usepackage{amsmath}

\begin{document}
\title{Rashba billiards}
\author{Andr\'as Csord\'as}
%\email{csordas@tristan.elte.hu}
\affiliation{Research Group for Statistical Physics of the
             Hungarian Academy of Sciences,
             P\'azm\'any P. S\'et\'any 1/A, H-1117 Budapest, Hungary}

\author{J\'ozsef Cserti}
%\email{cserti@complex.elte.hu}
\affiliation{Department of Physics of Complex Systems, E{\"o}tv{\"o}s
University, H-1117 Budapest, P\'azm\'any P{\'e}ter s{\'e}t\'any 1/A, Hungary}

\author{Andr\'as P\'alyi}
%\email{palyi@complex.elte.hu}
\affiliation{Department of Physics of Complex Systems, E{\"o}tv{\"o}s
University, H-1117 Budapest, P\'azm\'any P{\'e}ter s{\'e}t\'any 1/A, Hungary}

\author{Ulrich Z\"ulicke}
%\email{u.zuelicke@massey.ac.nz}
\affiliation{Institute of Fundamental Sciences and MacDiarmid Institute for
Advanced Materials and Nanotechnology, Massey University, Private Bag
11~222, Palmerston North, New Zealand}  

\date{\today}

\begin{abstract}
We study the energy levels of non-interacting electrons confined to move in
two-dimensional billiard regions and having a spin-dependent dynamics due
to a finite Rashba spin splitting. The Green's function for such Rashba
billiards is constructed analytically and used to find the area and perimeter
contributions to the density of states, as well as the smooth counting function. 
We show that, in contrast to systems with spin-rotational invariance, Rashba
billiards always possess a negative energy spectrum. 
A semi-classical analysis is presented to interpret the singular
behavior of the density of states at certain negative energies.  
Our detailed analysis of the spin structure of Rashba billiards reveals
a finite out-of-plane spin projection for electron eigenstates.
\end{abstract}


\maketitle

\section{Introduction}

Spin-dependent phenomena in semiconductor nanostructures have attracted
great current interest~\cite{spintronics-book,Wolf-review:cikk}. Intriguing
effects can arise in non-magnetic systems due to the presence of
spin--orbit coupling. Structural inversion asymmetry in semiconductor
heterostructures has been shown~\cite{roessler} to give rise to a spin
splitting of the same type as was discussed in an early paper by Rashba~\cite
{Rashba:cikk}. Its tunability by external gate voltages~\cite{nitta,thomas,syoji:jap:01}
has motivated the theoretical design of a spin--controlled field--effect
transistor~\cite{Datta-Das:cikk}. Novel spin properties arise from the interplay
between Rashba spin splitting and further confinement of two--dimensional
electrons in quantum wires~\cite{hausler,Mireles-Kirczenow:cikk,
Uli-1:cikk,thomas_wire}, rings~\cite{Uli_persistent:cikk,Benedict-1:cikk}, or 
dots~\cite{dot1,Michele-q-dot:cikk,dot2,Bulgakov-Sadreev:cikk,dot3,zaitsev,
Rodriguez}.  Spin-orbit coupling has also been shown to affect the statistics of
energy levels and eigenfunctions as well as current
distributions~\cite{Berggren:cikk,Sadreev:cikk}.  The interplay between spin-orbit
coupling and external magnetic fields was analyzed theoretically using
random matrix theory~\cite{falko}.

In this work, we study {\em Rashba billiards}, i.e., non-interacting
ballistic electrons moving in finite two-dimensional (2D) regions 
whose dynamics is affected by the Rashba spin--orbit coupling. 
In the one-band effective-mass approximation, the Hamiltonian with Rashba
splitting in 2D is given by~\cite{bychkov}
\begin{subequations}
\begin{eqnarray}
\hat{H} &=& \hat{H}_0 + \frac{\alpha}{\hbar}\,
\hat{U},  \\
\hat{H}_0 &=& \frac{p_x^2 + p_y^2}{2m^*}
\\  
\hat{U} &=& \sigma_x p_y - \sigma_y p_x \quad ,
\end{eqnarray}
\label{Rashba-Hamiltonian:eq}%
\end{subequations}%
where $\sigma_x,\sigma_y$ are Pauli matrices. 
This Hamiltonian governs
the electron dynamics inside the billiard with 
Dirichlet boundary conditions at the perimeter 
(see Ref.~\onlinecite{Bulgakov-Sadreev:cikk}).  
The Rashba spin--orbit coupling strength $\alpha$ can be conveniently 
measured in terms of a wave--number scale $k_{\text{so}} =m^*\alpha/\hbar^2$,
The spin--precession length defined as $\pi/k_{\text{so}}$ 
can be tuned independently of the system 
size~\cite{nitta,thomas,syoji:jap:01}.
Furthermore, the tunability of the Rashba spin--orbit
coupling strength is a convenient tool to induce changes of the billiard's
energy spectrum without applying external magnetic fields.

One of our central quantity of interest is the density of states (DOS) 
$\varrho(E)= \sum_n \delta (E-E_n)$ and the counting function $N(E) =
\sum_n \Theta (E-E_n)$ for Rashba billiards with energy levels $E_n$. 
Here $\delta(x)$ and $\Theta(x)$ are the Dirac delta function and 
the Heaviside  function, respectively.    
The density of states and the counting function of normal billiards 
(without spin--orbit coupling, i.e., for $\alpha =0$) 
have been extensively studied in the literature.
They can be derived from the Green's function of the system.
The smooth counting function $\bar{N}(E)$ is given by the so-called 
Weyl formula~\cite{Weyl:cikk,Baltes-Hilf:book,Brack:konyv}, which is 
an asymptotic series of the exact counting function $N(E)$, 
in terms of the energy $E$. 
A good introduction to this problem is given by Baltes and 
Hilf~\cite{Baltes-Hilf:book}, 
and recently by Brack and Bhaduri~\cite{Brack:konyv}, 
and many applications can be found in
Refs.~\onlinecite{Kac:cikk,Balian-Bloch:cikkek,Stewartson-Waechter:cikk,
Berry-Howls:cikk,Uzy-Sieber:cikk,Q-billiard:konyv}.

We have derived the first two leading terms in the Weyl formula of 
the smooth counting function $\bar{N}(E)$  
for arbitrary shapes of Rashba billiards.
Our approach is based on the image method of Berry and
Mondragon~\cite{Berry-neutrino:cikk} developed for neutrino billiards, 
which have two--component wave functions 
and in this respect are rather similar 
to the Rashba billiards discussed here.
We will show that the first term of $\bar{N}(E)$ is proportional to the
area of the billiard, while the second one is proportional to the
length of the perimeter of the billiard. 
Moreover, we find that the density of states is singular at the bottom of
the spectrum.
This singular behavior occurs independently of the billiard's shape and is
most striking if the Rashba parameter is large.

The circular Rashba billiards is the simplest of confined systems that
can be treated
analytically~\cite{Bulgakov-seq_eq,magneses-dot_SO-exact:cikk,malshukov}. 
Following the approach outlined above, we also calculate the 
smooth counting function $\bar{N}(E)$ for circular Rashba billiards,
and besides its first two leading terms (which coincide with the
results derived for arbitrary shapes of Rashba billiards) 
we give higher-order correction terms.
Our analytical result for $\bar{N}(E)$ and that obtained 
from the numerically calculated exact energy levels are  
in perfect agreement.

In the absence of any lateral confinement,
the energy dispersion for the Rashba 
Hamiltonian~(\ref{Rashba-Hamiltonian:eq})  
splits into two branches~\cite{bychkov}:
\begin{equation}
E (k_x,k_y) = \frac{\hbar^2}{2m^*}\, 
\left[{\left(|{\bf k}| \pm  k_{\text{so}}\right)}^2 -  k_{\text{so}}^2\right], 
\label{dispersion:eq}
\end{equation}
where ${\bf k} = (k_x,k_y)$. 
The spin splitting is a consequence of broken spin-rotational invariance. 
The spin of energy eigenstates, which are labeled by a 2D vector 
${\bf k}$, is polarized perpendicularly to ${\bf k}$~\cite{bychkov}. 
Hence, no common spin quantization axis for single--electron states
can be defined in the presence of spin--orbit coupling.
As can be seen, in the range $0< k< 2 k_{\text{so}}$, one branch
has {\em negative energies} bounded from below 
by $-\Delta_{\text{so}}\equiv -\hbar^2 k_{\text{so}}^2/(2m^*)$.

Similarly, a laterally confined 2D system in the presence of Rashba type 
spin-orbit interactions has also a negative energy spectrum. 
In this paper, we present interesting features of the energy spectrum for
circular Rashba billiards, focusing especially 
on its negative energy eigenvalues.
We have found that for a circular shape, the density of states has additional 
singularities at negative energies. 
We obtain analytic results for their positions. 
Their corresponding eigenspinors have a finite spin projection 
in the direction perpendicular to the billiard plane, which is the
direct result of imposing hard--wall boundary conditions.

Results presented in this article extend work reported in
Ref.~\onlinecite{JAU:paper}. Its organisation is as follows.
The properties of arbitrarily shaped Rashba billiards are discussed in
Sec.~\ref{arbitrary:ch}. We present an algebraic expression for the free-space
Green's function  in the presence of Rashba spin-orbit coupling 
in Sec.~\ref{free_space_G:ch}. Susbequently, in 
subsection~\ref{area_perim_DOS:ch}, the first two leading terms of 
the Weyl formula are derived. 
Using the eigenstates presented in Sec.~\ref{eigen_free_space:ch} 
in the absence of lateral confinement, we derive an alternative expression
for the free-space Green's function in polar coordinates  in
Sec.~\ref{freeG_pol:ch}. 
The circular Rashba billiards are discussed in Sec.~\ref{disk:ch}. 
An analytical formula for the Green's function is derived in
Sec.~\ref{disk_pos_G:ch} for this case, while 
the derivation of the smooth counting function is presented 
in Sec.~\ref{smooth_circ:ch}, including its comparison with the
numerically calculated result.
For negative energies the counting function is calculated in 
Sec.~\ref{negative_circ:ch}, while the spin structures is discussed in 
Sec.~\ref{spin_circ:ch}. 
Finally, our results are summarized and conclusions given
in Sec.~\ref{conclusion:ch}.

\section{Arbitrary shapes of Rashba billiards}
\label{arbitrary:ch}

In this section we derive the smooth part of the density of states and 
the smooth part of the counting function, i.e., the two leading terms 
in the Weyl formula~\cite{Weyl:cikk,Baltes-Hilf:book,Brack:konyv} 
for arbitrary shapes of Rashba billiards.  
These smooth functions are obtained by averaging the exact DOS and 
counting function over a small energy range around an energy $E$.   

The exact density of states $\varrho(E)$ expressed 
in terms of the retarded Green's function 
(see e.g.,~\cite{Brack:konyv}) is given by  
\begin{equation}
\varrho(E)=-\frac{1}{\pi}
\lim_{\eta \to 0^+}{\rm Im}\, {\rm Tr}\, G(E+i\eta,\mathbf{r},\mathbf{r}'),
\label{DOS:def}
\end{equation}
where the trace means the limit ${\bf r} \to {\bf r}^\prime$, 
integration of ${\bf r}$ over the area of the billiard, 
and the trace in spin space. 
The  exact Green's function $ G(z,\mathbf{r},\mathbf{r}')$ 
is the position representation of the 
Green operator $\hat{G}(z) = {(z-\hat{H})}^{-1}$, which in addition,
satisfies the boundary conditions.  
Then, the exact counting function is defined by $N(E) = \int_{-\infty}^E \, 
\varrho(E^\prime) d E^\prime$. 

Usually, the exact Green's function satisfying the boundary conditions
is not known. 
However, one can always write the exact Green's function as a sum of the 
so-called free-space Green's function and a correction with which the 
exact Green's function satisfies the boundary conditions. 
The free-space Green's function $G_\infty(E,\mathbf{r},\mathbf{r}')$ 
is the Green's function of the infinite system and 
does not satisfy the boundary conditions at the boundary of the billiards. 
In this paper, we calculate the free-space 
Green's function $G_\infty(E,\mathbf{r},\mathbf{r}')$ for the Rashba
Hamiltonian (\ref{Rashba-Hamiltonian:eq}), and 
in case of circular Rashba billiards, the exact Green's function which
satisfies the Dirichlet boundary conditions at the boundary of the billiards. 

The first term in the Weyl formula, called area term, can be obtained 
by replacing the exact Green's function with the free-space Green's
function $G_\infty(E,\mathbf{r},\mathbf{r}')$ in Eq.~(\ref{DOS:def}).
It is always proportional to the area of the billiard. 
Higher-order terms in the Weyl formula are the corrections to the
area term taking into account the exact Green's function. 
The smooth part of the first correction term is called perimeter term
because it is proportional to the length of the perimeter of the billiard.   

\subsection{Free-space Green's function for Rashba billiards}
\label{free_space_G:ch}

All our subsequent calculations are crucially based on the knowledge
of the  free-space Green's function $G_\infty(E,\mathbf{r},\mathbf{r}')$ for
Rashba billiards. In this subsection, we we present its derivation.   

At a given energy $E$, two propagating modes exist whose wave vectors 
can be found from the dispersion relation (\ref{dispersion:eq}): 
\begin{equation}
|{\bf k}| = k_{\pm} =|k \mp k_{\text{so}}|, \quad \text{where} \quad k=\sqrt
{\frac{2m^* E}{\hbar^2}+k_{\text{so}}^2}. 
\label{kpm:def}
\end{equation}

Using the identities for the Pauli matrices one can easily show 
that $\hat{U}^2 = p_x^2 +p_y^2$ and the Rashba Hamiltonian can be
written as 
$$\hat{H} = \frac{\hat{U}^2}{2m^*}+ \frac{\alpha}{\hbar}\,\hat{U} \quad .$$
The free-space Green operator $\hat{G}_\infty(E) = {(E-\hat{H})}^{-1}$ 
corresponding to the Rashba Hamiltonian reads then
\begin{equation}
\hat{G}_\infty(E) = \frac{2 m^*}{\hbar^2} 
{\Biggl[k^2(E) - {\left(\frac{\hat{U}}{\hbar} +  
k_{\text{so}} \right)}^2 \Biggr]}^{-1},  
\end{equation} 
where $k(E)$ is given by Eq.~(\ref{kpm:def}).
Here $E$ can be a complex number.
Using the operator identity 
\begin{equation}
{\left(\lambda^2 - \hat{A}^2\right)}^{-1}
= \frac{1}{2 \lambda}\, 
\Biggl[ 
{\left(\lambda + \hat{A} \right)}^{-1} 
+ {\left(\lambda - \hat{A} \right)}^{-1}
\Biggr ],
\end{equation}
where $\lambda$ is a scalar and $\hat{A}$ is an operator, 
one can decompose $\hat{G}_\infty(E)$ as 
\begin{equation}
\hat{G}_\infty(E) = \frac{m^*}{k\hbar^2} 
\Biggl[\!\! {\left(\! k_- + \frac{\hat{U}}{\hbar} \! \right)}^{-1} 
\!\!\! + \!\!
 {\left(\! \text{sgn}(E) k_+ - \frac{\hat{U}}{\hbar}\! \right)}^{-1} 
\Biggr] ,
\end{equation}
where $k_\pm$ are given by Eq.~(\ref{kpm:def}). 
Now using the operator identity 
${\left(\lambda \pm \hat{A}\right)}^{-1} = 
\left(\lambda \mp \hat{A}\right) {\left(\lambda^2 -\hat{A}^2\right)}^{-1}$, 
one finds
\begin{eqnarray}
\hat{G}_{\infty}(E) &=& 
\frac{m^*}{\hbar^2} \, 
\frac{1}{k} \, \left[
\left(k_- - \frac{\hat{U}}{\hbar}\right)
{\left(k^2_- -\frac{{\bf p}^2}{\hbar^2}\right)}^{-1}
\right.  \nonumber \\ 
&& \left. + \left( \text{sgn}(E)k_+ + \frac{\hat{U}}{\hbar} \right)
{\left(k^2_+ -\frac{{\bf p}^2}{\hbar^2}\right)}^{-1}   
\right] .
\label{free-Green:eq}
\end{eqnarray}

The retarded Green's function in position representation is given by 
\begin{equation}
G_\infty(E,\mathbf{r},\mathbf{r}')=
\langle \mathbf{r} \mid \hat{G}_\infty(E+i \eta) \mid \mathbf{r}'\rangle ,
\label{retG:def}
\end{equation}  
where $E$ is a real number and $\eta \to 0^+$.
The two terms in Eq.~(\ref{free-Green:eq}) in position representation 
involve two functions: 
\begin{eqnarray}
 \langle \mathbf{r} \mid  
{\left(k^2_\pm -\frac{{\bf p}^2}{\hbar^2}\right)}^{-1}
\mid \mathbf{r}'\rangle . 
\label{pos_rep_two-terms:eq}
\end{eqnarray}
After a simple limiting procedure one can show that 
\begin{subequations}
\begin{eqnarray}
k_+^2(E+i \eta) &=& k_+^2(E)+ \text{sgn}(E)\,  i \eta,
 \\[2ex]
k_-^2(E+i \eta) &=&  
k_-^2(E)+ i \eta .
\end{eqnarray}%
\label{k_limit:eq}%
\end{subequations}%
The two functions in (\ref{pos_rep_two-terms:eq}) 
can be evaluated by the following
identities (see e.g.,~\cite{Brack:konyv}): 
\begin{equation}
 \langle \mathbf{r} |  
{\left( k^2-\frac{\hat{{\bf p}}^2}{\hbar^2} \pm i \eta \right)}^{-1} 
\!\! |  \mathbf{r}'\rangle =
\!\! \begin{cases}   
-{i \over 4}H_0^{(1)}(k|\mathbf{r}-\mathbf{r}'|),  \\[2ex]
{i \over 4}H_0^{(2)}(k|\mathbf{r}-\mathbf{r}'|),
\end{cases}
\label{scalar_G:eq}
\end{equation}
where $H_0^{(1,2)}(x)$ are the Hankel functions of zero order, and $k>0 $.

Finally, using Eqs.~(\ref{free-Green:eq}) - (\ref{scalar_G:eq}) 
we can easily find
\begin{widetext}
\begin{equation} 
G_\infty(E,\mathbf{r},\mathbf{r}') =  
\frac{-i}{4} \frac{m^*}{\hbar^2} 
\frac{1}{k}  
\begin{cases}   
\left[ \left(k_- - \frac{\hat{U}}{\hbar} \right) 
H_0^{(1)}(k_-|\mathbf{r}-\mathbf{r}'|)
+\left(k_+ + \frac{\hat{U}}{\hbar} \right) 
H_0^{(1)}(k_+|\mathbf{r}-\mathbf{r}'|)
\right], \quad \text{for} \,\,  E > 0, \\[3ex]
\left[ \left(k_- - \frac{\hat{U}}{\hbar} \right) 
H_0^{(1)}(k_-|\mathbf{r}-\mathbf{r}'|)
-\left(- k_+ + \frac{\hat{U}}{\hbar} \right) 
H_0^{(2)}(k_+|\mathbf{r}-\mathbf{r}'|)
\right], \,\,  \text{for} \,\,  E < 0. 
\end{cases}
\label{eq:freegreen_fn}
\end{equation}
\end{widetext}
We note that, for negative energies $E$, the retarded Green's function 
contains incoming circular waves besides outgoing waves.

\subsection{Area and perimeter terms of the density of states}
\label{area_perim_DOS:ch}

First, consider the area term of the Weyl formula.
Now, in Eq. (\ref{DOS:def}) we replace  the exact Green's function 
$G(E,\mathbf{r},\mathbf{r}')$ by the free-space Green's function 
$G_\infty(E,\mathbf{r},\mathbf{r}')$ given by Eq.~(\ref{eq:freegreen_fn}). 
The trace of the operator $\hat{U}$ is zero 
since $\hat{U}$ is an off-diagonal matrix in the spin space. 
Then, it is easy to see that the leading term in the DOS becomes
\begin{equation} 
\varrho_{\text{area}} (E) = 
\frac{\cal{A}}{2\pi}\, \frac{2m^*}{\hbar^2}\,\frac{1}{k} \,
\begin{cases}
\begin{array}{c}  
k, \quad \text{for} \,\,  E > 0, \\[1ex]
k_{\text{so}}, \quad \text{for} \,\,  E < 0, 
\end{array}
\end{cases}
\label{leading_DOS:eq}
\end{equation} 
where $\cal{A}$ is the area of the Rashba billiard. 
Therefore, the integration of the DOS yields the counting function: 
\begin{equation}
N_{\text{area}}(E) =
\frac{{\cal A}}{\pi } \, \frac{2 m^*}{\hbar^2} \! \left\{
\begin{array}{l} \!
\frac{E}{2} + \Delta_{\text{so}}, \,\, \text{for}\,\, E> 0, \\[2ex]
\!\!\! \sqrt{\Delta_{\text{so}}}\sqrt{E+\Delta_{\text{so}}}, 
\, \text{for}\,-\Delta_{\text{so}}\!\!< \! E \! < 0. 
\end{array} \right.  
\label{leading_N:eq}
\end{equation}
It follows directly from Eq.~(\ref{leading_N:eq}) that, for negative energies,
the DOS shows a $1/\sqrt{E+\Delta_{\text{so}}}$ singularity at the bottom
of the spectrum $E \to -\Delta_{\text{so}}$.
The area term (\ref{leading_N:eq}) can alternatively be derived from
the classical phase-space integral in the underlying classical approach.
However, the classical dynamics of electrons in Rashba billiards is described
by {\em two} Hamiltonians \cite{Pletyukhov-Brack:cikk}, which  are reminiscent
of the two dispersion branches (\ref{dispersion:eq}). 
The constant-energy surfaces in phase space are different 
for the two Hamiltonians, yielding different contributions to 
the classical phase-space integral. 
This simple calculation also leads to Eq.~(\ref{leading_N:eq}). 

For arbitrary shapes of Rashba billiards, we can also determine 
the perimeter term of the DOS and the counting function.
This term can be derived from the generalization of the image 
method of Ref.~\onlinecite{Balian-Bloch:cikkek} using only the free space
Green's function. 
The actual calculation is very much similar to that applied by 
Berry and Mondragon~\cite{Berry-neutrino:cikk} for neutrino billiards.  
The Dirichlet boundary conditions can be approximately satisfied by regarding
the boundary as straight and using the approximate Green's function 
\begin{subequations}
\begin{eqnarray}
G(\mathbf{r},\mathbf{r}') & \approx & 
G_\infty(\mathbf{r},\mathbf{r}') +
G_i(\mathbf{r},\mathbf{r}'), \, \text{where} \, \\
G_i(\mathbf{r},\mathbf{r}') & \equiv & 
-G_\infty(\mathbf{r},\mathbf{r}_i),
\label{imageG:eq}
\end{eqnarray}%
\end{subequations}%
and $\mathbf{r}_i$ is the image of $\mathbf{r}'$ on the boundary
outside the billiard. 
Obviously $G(\mathbf{r},\mathbf{r}')$ is still a solution of the
Schr\"odinger equation in the variable $\mathbf{r}$. 
To calculate the trace in (\ref{DOS:def}) of 
$G_i(\mathbf{r},\mathbf{r}')$ we define  
${\bf r} = (n,s)$ and  ${\bf r}' = (n,s+\sigma)$, where 
$n$ and $s$ are the coordinates of ${\bf r}$ perpendicular to and along 
the boundary. 
Of course $n<0$ since ${\bf r}$ is inside the billiard and 
the limit ${\bf r} \to {\bf r}^\prime$ in the trace 
corresponds to $\sigma \to 0$.
Now, $\mid {\bf r} - \mathbf{r}_i \mid = 
\sqrt{{\left( 2n \right)}^2 + \sigma^2}$ and the correction to the DOS, 
i.e., $\varrho_{\text{perim}}  (E)$ coming from 
$G_i(\mathbf{r},\mathbf{r}')$ can be written as
\begin{eqnarray}
\varrho_{\text{perim}} (E) &=& -\frac{1}{\pi}
\lim_{\eta \to 0^+}{\rm Im}\, {\rm Tr}\,\, 
\hat{G_i}(E+i\eta,\mathbf{r},\mathbf{r}') \nonumber \\ 
&=& -\frac{1}{2\pi} \frac{m^*}{\hbar^2} \frac{1}{k} 
\int_0^{\cal L}\, ds \int_{-\infty}^0 \, dn \, 
\nonumber \\ 
&& \left[ k_- J_0(k_- 2n) + k_+ J_0\left(k_+ 2n \right) \right],
\end{eqnarray}
where the factor $2$ of the trace in the spin space has already been 
included.   
Using the integral $\int_0^\infty J_0(a x) dx = 1/a$ with $a>0$
we obtain
\begin{equation}
\varrho_{\text{perim}} (E) =
-\frac{\cal L}{4\pi}\frac{2m^*}{\hbar^2} \frac{1}{k},
\label{DOS_perim_gen:eq}
\end{equation}
valid for all energies $E>-\Delta_{\text{so}}$. 
Here $\cal{L}$ is the length of the perimeter of the billiard. 
Finally, the integration of the DOS yields the counting function: 
\begin{equation}
N_{\text{perim}}(E) = 
-\frac{\cal{L}}{2\pi}\, \sqrt{\frac{2m^*}{\hbar^2}}\, 
\sqrt{E+ \Delta_{\text{so}}},
\label{perim_N:eq}
\end{equation}
valid also for all energies $E>-\Delta_{\text{so}}$. 
The minus sign is a consequence of Dirichlet boundary conditions. 

In summary, the first two terms in the Weyl formula for arbitrary shapes
of Rashba billiards reads as 
$\bar{N} (E) = N_{\text{area}}(E)+N_{\text{perim}}(E)$.
Note that for zero spin--orbit coupling, 
$\bar{N} (E)$ coincides with the previously derived result for 2D 
billiards~\cite{Baltes-Hilf:book,Brack:konyv,Kac:cikk,Stewartson-Waechter:cikk,Berry-Howls:cikk} 
(apart from a factor 2 due to spin).

\subsection{Eigenstates for infinite systems}
\label{eigen_free_space:ch}
In this section the eigenvalues and eigenstates of the 
free-particle Rashba Hamiltonian given by
Eq.~(\ref{Rashba-Hamiltonian:eq}) are calculated in polar
coordinates. 
These results will be used in the subsection \ref{freeG_pol:ch} 
to rewrite the free-space Green's function (\ref{eq:freegreen_fn}) 
in a form which is suitable for calculations 
in case of circular Rashba billiards presented in Sec.~\ref{disk:ch}. 

The Hamiltonian (\ref{Rashba-Hamiltonian:eq}) 
can be rewritten in polar coordinates ${\bf r}=(r,\varphi)$ and we have 
$\hat{H} = \hat{H}_0 + \frac{\alpha}{\hbar}\, \hat{U}$, where   
\begin{subequations}
\begin{eqnarray}
\hat{H}_0  \!\! &=&  -\frac{\hbar^2}{ 2m^*}
\left( \frac{\partial^2}{\partial r^2}+\frac{1}{ r}
\frac{\partial}{\partial r}
+\frac{1}{r^2}\frac{\partial^2}{\partial \varphi^2} \right), \\[2ex]
\!\! \frac{\hat{U}}{\hbar} &=&   \!\!\!\! 
\left(\begin{array}{cc}
 \!\! 0 & e^{-i\varphi} \left(\frac{\partial}{\partial r}-\frac{i}{r}
\frac{\partial}{\partial \varphi}\right) \\
 \!\! - e^{i\varphi} \left(\frac{\partial}{\partial r}+\frac{i}{r}
\frac{\partial}{\partial \varphi}\right) & 0
\end{array} \!\!\! \right)   \! .
\label{U:def}
\end{eqnarray}%
\end{subequations}%
Since the  Hamiltonian $\hat{H}$ commutes with the total angular
momentum operator
$\hat{J}_z =-i\hbar \partial_\varphi + \frac{\hbar}{2}\,\sigma_z$, 
the stationary Schr\"odinger equation 
$\hat{H} \mid \chi \rangle = E \mid \chi \rangle $ can be
solved using the following ansatz~\cite{Bulgakov-Sadreev:cikk,Rodriguez} 
\begin{equation}
\langle  \, {\bf r} \, \mid \,\chi_m \rangle = 
\left(\begin{array}{c}
C_1 Z_m(kr)e^{i m \varphi}\\[1ex]
C_2 Z_{m+1}(kr)e^{i (m+1) \varphi}
\end{array}\right),
\end{equation} 
where $m$ is an integer, $k>0$ and $Z_m(x)$ can be any of the 
Bessel functions $J_m(x)$, $Y_m(x)$, and $H^{(1,2)}_m(x)$. 
With the help of the well-known recursion relations of Bessel
functions 
\begin{eqnarray}
Z_m'(x) \pm \frac{m}{x}\, Z_{m}(x) &=& \pm Z_{m\mp 1}(x),  
\label{eq:bess_rec}
\end{eqnarray}
one can show that the constants $C_1$ and $C_2$ satisfy 
\begin{equation}
\left(\begin{array}{cc}
k^2 & 2 k k_{so}\\[1ex]
2 k k_{so} & k^2   
\end{array}\right)
\left(\begin{array}{c}
C_1 \\[1ex] C_2  
\end{array}\right)= \frac{2 m^* E}{\hbar^2}\,  \left(\begin{array}{c}
C_1 \\[1ex] C_2 
\end{array}\right). 
\label{eq:cek}
\end{equation}   
Hence for a given $k$ the two eigenenergies $E_\pm$ are 
\begin{equation}
E_\pm (k) =  \frac{\hbar^2}{2m^*}\, 
\left[{\left( k\pm  k_{\text{so}}\right)}^2 - k_{\text{so}}^2\right]. 
\label{E_k:eq}
\end{equation}
Since the eigenvalues of the Schr\"odinger equation are independent of 
the chosen coordinate systems the above eigenenergies 
should be the same as those given in Eq.~(\ref{dispersion:eq}), 
which is indeed the case when $k = |{\bf k}|.$  
The corresponding two non-trivial solutions for $C_1^\pm$ and $C_2^\pm$
are given by
\begin{equation}
\begin{array}{l}
C_1^\pm / C_2^\pm = \pm 1,  \quad \mbox{for } E>0, \\[2ex]
C_1^\pm / C_2^\pm = -1,  \quad \mbox{for} -\Delta_{\text{so}} <E<0. 
\end{array}
\end{equation}
For a given $E$ the two positive 
solutions of Eq.~(\ref{E_k:eq}) for $k$ are $k_\pm$ given 
by Eq.~(\ref{kpm:def}). 

We are now in a position to construct different eigenstates using the 
Bessel and Hankel functions.  
The eigenspinors regular at the origin are 
\begin{equation}
\langle  \, {\bf r} \mid \chi_m^{\pm} \rangle =
\begin{cases}
\left(\begin{array}{c}
\pm J_m(k_\pm r)\\[1ex]
J_{m+1}(k_\pm r)e^{i\varphi}
\end{array}\right)e^{im\varphi},\, E>0,
  \\[4ex]
\left(\begin{array}{c}
- J_m(k_\pm r)\\[1ex]
J_{m+1}(k_\pm r)e^{i\varphi}
\end{array}\right)e^{im\varphi},\, E<0.
\end{cases}
\label{eq:chi} 
\end{equation}
To derive the free-space Green's function in polar coordinates 
we shall also use solutions which are singular at the origin:
\begin{subequations}
\begin{equation}
\langle  \, {\bf r} \mid h_m^{\pm} \rangle =
\left(\begin{array}{c}
\pm H_m^{(1)}(k_\pm r)\\[1ex]
H_{m+1}^{(1)}(k_\pm r)e^{i\varphi}
\end{array}\right)e^{im\varphi},\quad E>0,
\end{equation} 
\begin{equation}
\langle  \, {\bf r} \mid h_m^{+} \rangle =
\left(\begin{array}{c}
- H_m^{(2)}(k_+ r)\\[1ex]
H_{m+1}^{(2)}(k_+ r)e^{i\varphi}
\end{array}\right)e^{im\varphi},\quad E<0,
\end{equation} 
\begin{equation}
\langle  \, {\bf r} \mid h_m^{-} \rangle =
\left(\begin{array}{c}
- H_m^{(1)}(k_- r)\\[1ex]
H_{m+1}^{(1)}(k_- r)e^{i\varphi}
\end{array}\right)e^{im\varphi},\quad E<0.
\end{equation}% 
\label{eq:hm}%
\end{subequations}%

\subsection{Free-space Green's function in polar coordinates}
\label{freeG_pol:ch}

Using (\ref{eq:freegreen_fn}) and (\ref{U:def})  
the free-space retarded Green's function in the two energy ranges becomes 
\begin{widetext}
\begin{equation}
G_\infty(E,\mathbf{r},\mathbf{r}')=
-{im^* \over 4\hbar^2 k}
\begin{cases}
\left(\begin{array}{cc}
k_- H^1_- + k_+H^1_+&- e^{-i\varphi}
\left( {\partial \over \partial r}-{i\over r}
{\partial \over \partial \varphi}\right)(H^1_- -H^1_+) \\
e^{i\varphi}
\left( {\partial \over \partial r}+{i\over r}
{\partial \over \partial \varphi}\right)(H^1_- -H^1_+)& k_- H^1_- + k_+H^1_+\\
\end{array} \right),\quad E>0,
\\[5ex]
\left(\begin{array}{cc}
k_- H^1_- + k_+H^2_+&- e^{-i\varphi}
\left( {\partial \over \partial r}-{i\over r}
{\partial \over \partial \varphi}\right)(H^1_- +H^2_+) \\
e^{i\varphi}
\left( {\partial \over \partial r}+{i\over r}
{\partial \over \partial \varphi}\right)(H^1_- +H^1_+)& k_- H^1_- + k_+H^2_+\\
\end{array} \right),\,\, -\Delta_{\text{so}} < E < 0,  %
\end{cases}
\label{G_free_pol:eq}%
\end{equation}%
\end{widetext}%
where we used the notations 
$H_\pm^{1,2}\equiv H_0^{(1,2)} (k_\pm|\mathbf{r}-\mathbf{r}'|)$.
In the off-diagonal elements the differentiations with respect to $r$ 
and $\varphi$ can be carried out by introducing a new variable 
$\mbox{\boldmath $\rho$} =\mathbf{r}-\mathbf{r}'$. 
Then, for $E>0$ our simple algebraic method yields the same result 
that was derived by Walls et al.~\cite{Walls_Heller:cikk} 
using a different approach.
However, they do not present any explicit form for $E<0$.   

In our previous paper~\cite{JAU:paper} we used another form 
for the free-space Green's function (although as a lack of space 
it was not published there) in order to determine exactly 
the Green's function for circular Rashba billiards. 
In this approach differentiations with respect to $r$ and
$\varphi$ in the off-diagonal elements were performed using 
the addition theorem of the Bessel functions~\cite{Abramowitz-Stegun:konyv}
\begin{equation}
H_0^{(1,2)}(|\mathbf{r}-\mathbf{r}'|)=\sum_{m=-\infty}^\infty H_m^{(1,2)}(r)
J_m(r') e^{im(\varphi-\varphi')}, \quad r>r'
\end{equation}
and the recursion relations (\ref{eq:bess_rec}).
Then, for the free-space Green's function in polar
coordinates for $r>r^\prime$ we obtain a rather compact form 
in terms of the spinors defined in 
(\ref{eq:chi})--(\ref{eq:hm}): 
\begin{equation}
G_\infty (\mathbf{r}, \mathbf{r}') \! =
 c \!\!\! \sum_{m=-\infty}^\infty 
\!\! \Bigl[
k_+ \langle \mathbf{r} | h_m^+ \rangle 
\langle \chi_m^+ | \mathbf{r}' \rangle 
+ k_- \langle \mathbf{r} | h_m^- \rangle 
\langle \chi_m^- | \mathbf{r}' \rangle 
\Bigr],
\label{eq:greeninf}
\end{equation}
where %$c=- \frac{im^*}{4\hbar^2 k}$. 
$c=- im^* /(4\hbar^2 k)$. 
We shall use this form in subsection \ref{disk_pos_G:ch}.

\section{Circular Rashba billiards}
\label{disk:ch}

We now consider a circular Rashba billiard of radius $R$. 
The eigenstates of the system can be written 
as a linear combination of the regular eigenspinors given by 
Eq.~(\ref{eq:chi}) and the linear combination coefficients are chosen
such that the eigenstates satisfy the Dirichlet boundary conditions. 
The straightforward calculation yields the following secular equation: 
\begin{equation} 
J_m(k_+R)J_{m+1}(k_-R) + \text{sgn}(E) J_m(k_-R)J_{m+1}(k_+R)=0,
\label{sec_circ:eq}
\end{equation} 
where $m$ is an integer.
For each quantum number $m$ the solutions of this equation for $E$
give the energy levels of the circular Rashba billiards.
The same secular equation was derived 
in Refs.~\cite{Bulgakov-seq_eq,magneses-dot_SO-exact:cikk,malshukov}. 
This equation is invariant under the change $m \rightarrow -m-1$ 
(Kramers degeneracy). 
Formal solutions of the secular equation having zero wave vectors
$k_+$ or $k_-$ are excluded since the corresponding wave functions 
vanish everywhere inside the billiard.
One such solution is at $E = -\Delta_{\text{so}}$. 

Following the ideas of the systematic method of 
Berry and Howls~\cite{Berry-Howls:cikk}, we have calculated the first 
few leading terms of the smooth counting function $\bar{N}(E)$. 
To do this we need the exact Green's function for circular Rashba
billiards which is calculated in the following subsection.

\subsection{Green's function for circular Rashba billiards}
\label{disk_pos_G:ch}

Boundary conditions for billiards requires that the Green's
function should vanish at the boundary (i.e., if either $\mathbf{r}$ or
$\mathbf{r}'$ is on the perimeter).
The free-space Green's function (\ref{eq:greeninf}) for a given energy 
usually does not vanish at the boundary of the billiard.
To fulfill the billiard boundary conditions we look
for the exact Green operator, as usual,  
in the form of $\hat{G}=\hat{G}_\infty + \hat{G}_H$,
where the homogeneous Green's function satisfies  
$\left(E- \hat{H} \right ) \hat{G}_{H} = 0$.
The boundary conditions for $\hat{G}$ are
\begin{equation}
G({\bf r}, {\bf r}^\prime) = G_{\infty} ({\bf r}, {\bf r}^\prime) 
+ G_{H} ({\bf r}, {\bf r}^\prime) = 0, 
\quad \text{for} \quad |{\bf r}|=R, 
\label{boundary:eq}
\end{equation}
where ${\bf r}^\prime$ is inside the billiard. 
Since the homogeneous Green's function $\hat{G}_H$ satisfies the same 
Schr\"odinger equation as the regular solutions given by
Eq.~(\ref{eq:chi}) one can construct $\hat{G}_H$ from these eigenstates as 
\begin{eqnarray}
\hat{G}_{H} &=& 
\sum_{m=-\infty}^\infty \Bigl[  \,\,  
A_m \, \mid \chi^+_m \rangle  \langle   \chi^+_m \mid 
+ B_m \, \mid  \chi^-_m \rangle  \langle   \chi^+_m \mid 
\Bigr.  \nonumber \\[1ex]
&& \Bigl. 
+C_m \, \mid \chi^+_m \rangle  \langle   \chi^-_m \mid 
+ D_m \, \mid \chi^-_m \rangle  \langle   \chi^-_m \mid 
 \,\, \Bigr],
\label{eq:greenhom}
\end{eqnarray}
where the constants $A_m, B_m, C_m$, and $D_m$, in
principle, can be determined from the boundary conditions 
(\ref{boundary:eq}). 
For arbitrary shapes of billiards it results in an infinite 
set of linear equations  for the constants. 
Fortunately, in case of circular billiards the constants 
can be determined analytically. 
Indeed, substituting Eqs.~(\ref{eq:greeninf}) and ~(\ref{eq:greenhom}) 
into Eq.~(\ref{boundary:eq}), and 
identifying the coefficients of the eigenspinors 
$\langle \chi_m^\pm \mid \mathbf{r}' \rangle$ one finds 
\begin{subequations}
\begin{eqnarray}
A_m \, \langle {\bf r} \mid \chi^+_m  \rangle 
+ B_m \, \langle {\bf r} \mid \chi^-_m \rangle \!\! &=& \!\! 
-c \, k_+ \, \langle {\bf r} \mid h^+_m \rangle , \\[2ex] 
C_m \, \langle {\bf r} \mid \chi^+_m \rangle 
+ D_m \, \langle {\bf r} \mid \chi^-_m \rangle \!\! &=& \!\! 
-c \, k_- \, \langle {\bf r} \mid h^-_m \rangle ,
\end{eqnarray}%
\label{ABCD:eq}%
\end{subequations}%
where the eigenspinors are evaluated at $|{\bf r}|=R$. 
These equations, in fact, are four independent linear inhomogeneous
equations for $A_m, B_m, C_m$, and $D_m$ 
since each eigenspinor is a two component vector. 
The solutions can be easily obtained from the
appropriate determinants formed from the coefficients of
Eq.~(\ref{ABCD:eq}), and are given by 
\begin{subequations}
\begin{equation}
A_m = -\frac{c \, k_+ }{ F_m} 
\begin{cases}
\left|\begin{array}{cc}
H_m^{(1)}(k_+R) & -J_m(k_-R)\\
H_{m+1}^{(1)}(k_+R) &J_{m+1}(k_-R) \\
\end{array}\right|, \, E>0,
\\[4ex]
\left|\begin{array}{cc}
-H_m^{(2)}(k_+R) & -J_m(k_-R)\\
H_{m+1}^{(2)}(k_+R) &J_{m+1}(k_-R) \\
\end{array}\right|,  \, E<0,
\end{cases}
\label{A:eq}
\end{equation}
\begin{equation}
D_m = -\frac{c \, k_- }{ F_m} 
\begin{cases}
\left|\begin{array}{cc}
J_m(k_+R) & -H_m^{(1)}(k_-R)\\
J_{m+1}(k_+R) & H_{m+1}^{(1)}(k_-R)\\
\end{array}\right|, \, E>0,
\\[4ex]
\left|\begin{array}{cc}
-J_m(k_+R) & -H_m^{(1)}(k_-R)\\
J_{m+1}(k_+R) & H_{m+1}^{(1)}(k_-R)\\
\end{array}\right|, \, E<0,
\end{cases}
\label{D:eq}
\end{equation}
\begin{equation}
B_m = C_m = \frac{2ic} {\pi R F_m}, \quad \text{where} 
\label{BC:eq}
\end{equation}
\begin{equation} 
F_m = J_m(k_+R)J_{m+1}(k_-R) + \text{sgn}(E) J_m(k_-R)J_{m+1}(k_+R).
\label{eq:F}
\end{equation}% 
\label{ABCDF:eq}% 
\end{subequations}%
In Eq.~(\ref{BC:eq}) we used the Wronskian relations for the Bessel
functions~\cite{Abramowitz-Stegun:konyv}. 

Finally, the analytical form of the exact retarded Green's function of 
circular Rashba billiards in polar coordinates is a sum of 
the free-space Green's function $G_{\infty} ({\bf r}, {\bf r}^\prime)$ 
given by Eq.~(\ref{G_free_pol:eq}) or Eq.~(\ref{eq:greeninf}), 
and the homogeneous part 
$G_H ({\bf r}, {\bf r}^\prime) = 
\langle {\bf r} \mid \hat{G}_H \mid {\bf r}^\prime \rangle$, 
where the operator $\hat{G}_H$ is given by 
Eq.~(\ref{eq:greenhom}) together with Eq.~(\ref{ABCDF:eq}). 

The eigenenergies of any billiards can be obtained from the poles of 
the retarded Green's function $\hat{G}$. 
For circular Rashba billiards the poles of $\hat{G}$ are the poles of 
$\hat{G}_H$, i.e.,  the zeros of $F_m$.
As can be seen it yields the same secular equation (\ref{sec_circ:eq}) 
derived independently, and thus it provides one check point for the Green's
function $\hat{G}_H$.

\subsection{The smooth part of the density of states }
\label{smooth_circ:ch}

To calculate the DOS and the counting function for circular Rashba
billiards we adopt the ideas of the systematic method of 
Berry and Howls~\cite{Berry-Howls:cikk}.
The exact Green operator of the system is 
$\hat{G} = \hat{G}_{\infty}+\hat{G}_H$. 
The first term of the density of states (\ref{DOS:def})  is the contribution 
from $\hat{G}_{\infty}$ in the trace of $\hat{G}$. 
The result is given in (\ref{leading_DOS:eq}), while 
the leading term in the counting function $N(E)$ is given by 
(\ref{leading_N:eq}). 

The correction terms of the DOS can be obtained from the trace of 
$\hat{G}_H$ given by Eq.~(\ref{eq:greenhom}). 
This involves the limit ${\bf r} \to {\bf r}'$, the trace of the 2 by 2 
matrix $G_H ({\bf r}, {\bf r})$ (trace over spinor indices) 
and the integration over the area of the billiard. 
After a straightforward calculation we found
\begin{eqnarray}
&&\textrm{Tr}\,\hat{G}_H = 2\pi\sum_{m=-\infty}^\infty \int_0^R
\, rdr \Biggl[
J_m^2(k_+r)(A_{m-1}+A_m) 
\Biggr. \nonumber \\
&& + J_m^2(k_-r)(D_{m-1}+D_m) 
\Biggl. \nonumber \\
&&  \Biggl. +2J_m(k_+r)J_m(k_-r) \left\{
\begin{array}{c}
\! (B_{m-1} - B_m), \,\, E > 0 \\[2ex] 
\! (B_{m-1} + B_m), \,\, E < 0 
\end{array} 
\right.
\Biggr].
\label{TrG_H:eq}
\end{eqnarray}
In the series with terms $A_{m-1}$ the summation index $m$ has been shifted 
by one to have the same radial integral as that in the series for $A_m$, 
and the same trick was done for series containing 
$B_{m-1}, C_{m-1}$ and $D_{m-1}$. 

The radial integrals in (\ref{TrG_H:eq}) can be performed
analytically~\cite{Gradshteyn-Ryzhik:konyv}. 
To calculate the density of states one needs to evaluate 
$\textrm{Tr}\,\hat{G}_H$ at complex energies $E+i\eta$.
To this end we follow the approach originally applied 
by Stewartson and Waechter\cite{Stewartson-Waechter:cikk}, 
and later for example Berry and Howls~\cite{Berry-Howls:cikk},  
and the Bessel functions of the first kind $J_m(z)$ and $H_0^{(1,2)}(z)$ 
are converted to the modified Bessel functions $I_m(z)$ and  $K_m(z)$
by extending the energy $E$ to the complex plane.
This is the so-called heat-kernel method.
However, in our case one has to be careful for negative energies. 
It turns out that the parameters $x$, $x_+$ and $x_-$ depending on 
energy $E$ (here $E$ is real) and defined as 
\begin{subequations}
\begin{eqnarray}
ix &\equiv& R \, k(E+i\eta),  \\
ix_+&\equiv&\textrm{sgn}(E) R \, k_+(E+i\eta) ,  \\
ix_-&\equiv& R \, k_-(E+i\eta), 
\label{eq:transformation}
\end{eqnarray}%
\end{subequations}%
are useful to convert 
the Bessel functions of the first kind 
to the modified Bessel functions using the identities  
\begin{subequations}
\begin{eqnarray}
J_m(iz) &=& i^m I_m(z),   \\
H_m^{(1)}(iz) \!\!\! &=& \!\!\! {2 \over \pi}(-i)^{n+1}K_m(z), \, 
-\pi \! < \! \arg{z}\le \! \frac{\pi}{2} \! , \\
H_m^{(2)}(-iz) \!\!\! &=& \!\!\! {2 \over \pi}i^{n+1}K_m(z),  \,  
-\frac{\pi}{2} < \arg{z}\le \pi.
\end{eqnarray}%
\end{subequations}%
After a tedious algebra the result of these transformations 
in Eq.~(\ref{TrG_H:eq}) can be written as 
\begin{subequations}
\begin{widetext}
\begin{eqnarray}
&& \textrm{Tr}\,\hat{G}_H(E+i\eta) = {m^* R^2 \over \hbar^2 x}
\sum_{m=-\infty}^\infty  f_m(E+i\eta), \quad \text{where} \nonumber \\[2ex] 
&& f_m(E+i\eta) = 
\Biggl[
1+{m^2 \over x_+^2} -\left({I_m'(x_+)\over I_m(x_+)} \right)^2
\Biggr]x_+ I_m(x_+) K_m(x_+)+
\Biggl[
1+{m^2 \over x_-^2} -\left({I_m'(x_-)\over I_m(x_-)} \right)^2
\Biggr]x_- I_m(x_-) K_m(x_-) \nonumber \\[3ex]
&& -\frac{P_m(x_+,x_-)}{2} \Biggl[
1+{m^2 \over x_+^2} -\left({I_m'(x_+)\over I_m(x_+)} \right)^2 
+ 1+{m^2 \over x_-^2} -\left({I_m'(x_-)\over I_m(x_-)} \right)^2
\!\!\! -\frac{4}{x_+^2-x_-^2}
\Biggl( x_+ {I_m'(x_+)\over I_m(x_+)} - x_- {I_m'(x_-)\over I_m(x_-)}\Biggr)
\Biggr] ,  
\\[2ex] 
&&  \text{and}  \nonumber  \\[2ex] 
&& P_m(x_+,x_-)=
{1 \over \displaystyle{{I_m'(x_+)\over I_m(x_+)} + {I_m'(x_-)\over I_m(x_-)}-
{m \over x_+ }-{m \over x_- }}}+
{1 \over \displaystyle{{I_m'(x_+)\over I_m(x_+)} + {I_m'(x_-)\over I_m(x_-)}+
{m \over x_+ }+{m \over x_- }}}.
\end{eqnarray}
\end{widetext}%
\label{f_m_E:eq} 
\end{subequations}% 
Up to now the trace of $G_H$ for circular Rashba billiards is exact. 
Note that the above mentioned transformations result in the same form
of $\textrm{Tr}\,\hat{G}_H(E+i\eta)$ both for negative and positive
energy $E$. 
Moreover, as a self-consistent check, one can show that $f_m(E)$ 
becomes the same as that 
in Refs.~\cite{Stewartson-Waechter:cikk,Berry-Howls:cikk}, 
when the spin-orbit coupling is zero, i.e., for $x_+ \to x_-$. 
Indeed, in this limit, using the L'Hospital's rule and the Bessel differential
equation for $I_m$, it can be shown that the factor multiplied 
by $P_m(x_+,x_-)$ in Eq.~(\ref{f_m_E:eq}) is exactly zero, 
and the remaining terms can be rewritten in the same form as that in
Refs.~\cite{Stewartson-Waechter:cikk,Berry-Howls:cikk}. 

The next step is to replace the modified Bessel functions in
Eq.~(\ref{f_m_E:eq}) by their uniform 
approximation~\cite{Abramowitz-Stegun:konyv,Berry-Howls:cikk}. 
Keeping only the leading terms we obtain  
\begin{eqnarray}
&& \textrm{Tr}\, \hat{G}_H = {m^* R^2 \over h^2 x}\sum_{m=-\infty}^\infty
\left\{ 
{1 \over 2}\left[{x_+ \over m^2 +x_+^2} +{x_- \over m^2 +x_-^2}\right]
\right. \nonumber \\
&& \left. +{1 \over x_+ +x_-}\left[ 1-{2m^2+x_+^2+x_-^2
\over 2\sqrt{(m^2 +x_+^2)(m^2 +x_-^2)}}\right]
\right\} . 
\label{leadingTrG_H:eq}   
\end{eqnarray}
Note that the second term bracketed in square brackets is zero when 
$x_+ \to x_-$, and one finds the perimeter term of the DOS for
billiards with zero spin-orbit coupling from the remaining terms. 
Taking into account the subsequent terms in the uniform approximation
provides a systematic way to derive higher order terms for the trace of
$G_H$ as in Ref.~\onlinecite{Berry-Howls:cikk} for normal circular billiards. 
However, with non-zero spin-orbit coupling the calculations become
more cumbersome. 

The summation over $m$ in~(\ref{leadingTrG_H:eq}) can be rewritten 
using the Poisson summation formula~\cite{Berry_Chaos:book,Brack:konyv} 
\begin{equation}
\sum_{m=-\infty}^\infty f_m 
= \sum_{\mu=-\infty}^\infty \int_{-\infty}^\infty d m f_m \, e^{i 2\pi \mu m}.
\label{Poisson_sum:eq} 
\end{equation}
Then, the Weyl series, i.e., the smooth part of the DOS, following 
Berry and Howls~\cite{Berry-Howls:cikk}, can be obtained by keeping
only the $\mu =0$-term in~(\ref{Poisson_sum:eq}). 
Carrying out the limiting process, $\eta \to 0$ in the trace of $G_H$
given by Eq.~(\ref{leadingTrG_H:eq}), and using the integral
\begin{eqnarray}
&& \int_b^a \, dz \, {2z^2-a^2-b^2\over\sqrt{(z^2-b^2)(a^2-z^2)}}
\nonumber \\
&& = 2(a+b) \left[ E\left({a-b \over a+b} \right)
- K\left({a-b \over a+b} \right)  \right],  
\end{eqnarray}
valid for $0<b<a$ ($E$ and $K$ are the complete elliptic integrals 
with the same definitions as in Ref.~\onlinecite{Gradshteyn-Ryzhik:konyv}) 
for integration over $m$, we obtain the contribution to 
the smooth DOS coming from $\textrm{Tr}\,\hat{G}_H$.   
A tedious calculation yields  
\begin{eqnarray}
&& \bar{\varrho}_H(\varepsilon) = 
-\frac{1}{2\sqrt{\varepsilon + \varepsilon_{\text{so}}}} 
- \, \sqrt{\varepsilon_{\text{so}}}\,
\delta(\varepsilon+ \varepsilon_{\text{so}})  
\nonumber \\
&& - \frac{1}{\pi} 
\begin{cases}
\frac{1}{\sqrt{\varepsilon + \varepsilon_{\text{so}}}} 
\Biggl[   
E\left(\! \sqrt{\frac{\varepsilon_{\text{so}}}
{\varepsilon + \varepsilon_{\text{so}}}} \, \right) 
- K\left(\! \sqrt{\frac{\varepsilon_{\text{so}}}
{\varepsilon + \varepsilon_{\text{so}}}} \, \right)
\Biggr]
, \, \varepsilon > 0,   \\[4ex]
\frac{\sqrt{\varepsilon_{\text{so}}}}
{\varepsilon + \varepsilon_{\text{so}}} \, 
\Biggl[
E\left(\! \sqrt{\frac{\varepsilon + \varepsilon_{\text{so}}}
{\varepsilon_{\text{so}}}} \, \right)
- K\left(\! \sqrt{\frac{\varepsilon + \varepsilon_{\text{so}}}
{\varepsilon_{\text{so}}}} \, \right) 
\Biggr],\,
\varepsilon <0 ,
\end{cases}%
\label{DOS_H:eq}
\end{eqnarray}%
where the dimensionless energies 
$\varepsilon = 2m^*ER^2/\hbar^2$ and 
$\varepsilon_{\text{so}} = 2m^* \Delta_{\text{so}} R^2/\hbar^2 
= k_{\text{so}}^2 R^2$ have been introduced.
The first term is the contribution from the first and second terms 
of Eq.~(\ref{leadingTrG_H:eq}), and it coincides with 
the perimeter term derived in Eq.~(\ref{DOS_perim_gen:eq}) 
for arbitrary shapes of Rashba billiards. 
The Dirac delta term and the terms containing the complete elliptic 
integrals in (\ref{DOS_H:eq}) come from the term 
involving bracets in (\ref{leadingTrG_H:eq}).

Finally, including the contribution from $\textrm{Tr}\,\hat{G}_\infty$, 
the integration of the DOS over $E$ leads to the smooth counting function
$\bar{N}(\varepsilon)$ for Rashba billiards: 
\begin{widetext}
\begin{equation}
\bar{N}(\varepsilon) = \left\{ \begin{array}{l}
\frac{\varepsilon+2\varepsilon_{\text{so}}}{2}
-\sqrt{\varepsilon+\varepsilon_{\text{so}}} 
+ \frac{2}{\pi} \Bigl[
\frac{\varepsilon}{\sqrt{\varepsilon + \varepsilon_{\text{so}}}}\, 
K\left(\! \sqrt{\frac{\varepsilon_{\text{so}}}
{\varepsilon + \varepsilon_{\text{so}}}} \, \right) 
- \sqrt{\varepsilon + \varepsilon_{\text{so}}} \, 
E\left(\! \sqrt{\frac{\varepsilon_{\text{so}}}
{\varepsilon + \varepsilon_{\text{so}}}} \, \right) 
\Bigr], 
\,\, \text{for}\,\, \varepsilon  > 0,  \\[4ex]
\sqrt{\varepsilon_{\text{so}}}\sqrt{\varepsilon + \varepsilon_{\text{so}}} 
-\sqrt{\varepsilon + \varepsilon_{\text{so}}} 
-\frac{2\sqrt{\varepsilon_{\text{so}}}}{\pi} \, 
E\left(\! \sqrt{\frac{\varepsilon + \varepsilon_{\text{so}}}
{\varepsilon_{\text{so}}}} \, \right), 
\,\, \text{for}\,\, -\varepsilon_{\text{so}} < \varepsilon  < 0.  
\end{array} \right.
\label{N_E-circular:eq}
\end{equation}
\end{widetext}
The first two terms (for both positive and negative energies) 
are the contribution from $\hat{G}_\infty$. 
They are the area and perimeter terms in the Weyl series and agree with 
the results given by (\ref{leading_N:eq}) and (\ref{perim_N:eq}), 
respectively for arbitrary shapes of Rashba billiards. 
The terms containing the complete elliptic integrals are corrections
to the perimeter term in case of circular billiards. 
We note that in a completely different context, namely for annular
ray-splitting billiards, a similar Weyl formula has been
calculated~\cite{ray-splitting:cikk} involving also
elliptic integrals.    

\begin{figure}[t]
\includegraphics[scale=0.57]{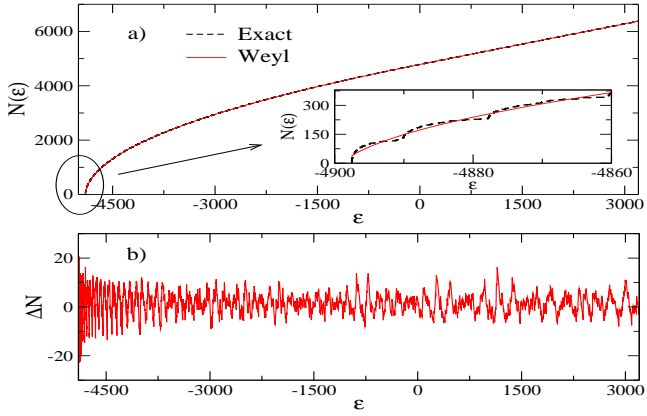}
\caption{(Color online) In panel a) the exact counting function 
$N(\varepsilon)$ (dashed line) and 
$\bar{N}(\varepsilon) $ (solid line) are shown for 
$\sqrt{\varepsilon_{\text{so}}}=k_{\text{so}} R =70$.
The inset shows the enlarged portion of the main figure close to the
bottom of the spectrum.
In panel b) the difference $\Delta N = N(\varepsilon)-\bar{N}(\varepsilon)$  
is plotted.
In both panels dimensionless energies $\varepsilon = 2m^*ER^2/\hbar^2$
are used.  
\label{N_E:figure}}
\end{figure}

We have compared the smooth counting function $\bar{N}(\varepsilon)$ 
given by Eq.~(\ref{N_E-circular:eq}) with the exact counting function 
$N(\varepsilon)$ calculated from the energy levels obtained from 
the secular equation (\ref{sec_circ:eq}) for different $m$.  
The relevant parameter characterizing a circular 
Rashba billiard of size $R$ is $k_{\text{so}} R$. 
Typical values for the spin--orbit--induced spin precession length
$L_{\text{so}}=\pi/k_{\text{so}}$ are of the order of a few hundred
nanometers~\cite{spintronics-book}. 
Taking $R=10 \mu$m for a typical size of quantum dots, 
the relevant parameter $k_{\text{so}} R$ in Rashba billiards 
can be as large as 70.
Figure~\ref{N_E:figure}a shows the exact and the smooth counting
functions as functions of the dimensionless energy $\varepsilon$.
There are 6388 energy levels in the plotted energy range. 
To see the difference between the two functions, in the inset 
we plotted them close to the bottom of the energy spectrum.   
Figure~\ref{N_E:figure}b shows the difference 
$\Delta N=N(\varepsilon)-\bar{N}(\varepsilon)$ as a function of $\varepsilon$.
The difference fluctuates around zero, which means we did not miss levels
(the mean value of $\Delta N$ is a sensitive test for missing levels, 
see e.g., Ref.~\onlinecite{Schmit:cikk_Csordas:cikk}). 
\begin{figure}[t]
\includegraphics[scale=0.35]{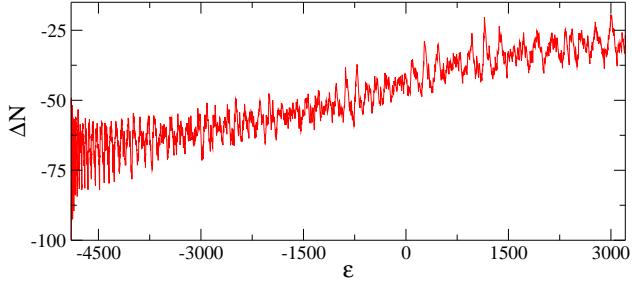}
\caption{The difference $\Delta N$ between the exact counting function 
and $\bar{N}(\varepsilon) $ without the terms containing the 
elliptic integrals in Eq.~(\ref{N_E-circular:eq}) is plotted.
The energy is scaled as $\varepsilon = 2m^*ER^2/\hbar^2$ 
and $k_{\text{so}} R =70$.  
\label{Delta_N_E-no_elliptic:figure}}
\end{figure} 
Without correction terms in Eq.~(\ref{N_E-circular:eq}) with elliptic
integrals, $\Delta N$ would increase monotonically on average 
as shown in Fig.~\ref{Delta_N_E-no_elliptic:figure}, and would
predict a difference $\approx 27$ in the energy range plotted.

\subsection{The counting function for negative energies}
\label{negative_circ:ch}

\begin{figure}[t]
\includegraphics[scale=0.35]{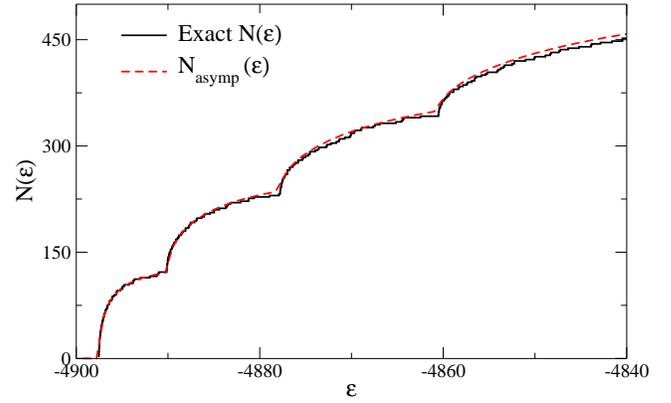}
\caption{(Color online) The exact counting function $N(\varepsilon)$ 
(solid line) and the asymptotic counting function 
$N_{\text{asymp}}(\varepsilon)$ 
given by Eq.~(\ref{N_asymp-2:eq}) (dashed line).
The energy is scaled as $\varepsilon = 2m^*ER^2/\hbar^2$ 
and $k_{\text{so}} R =70$.  
\label{N_E-rounded:figure}}
\end{figure}

In Fig.~\ref{N_E-rounded:figure}, the exact counting function 
is shown for negative energies near the bottom of the spectrum 
$ -\varepsilon_{\text{so}}$. 
As can be seen the exact $N(\varepsilon)$ shows an additional 
rounded step structure at certain energies $\varepsilon^*_n$. 
This feature shows up only for negative energies, although for 
larger energies this is less pronounced.  
The step structure results in large deviations $\Delta N$ at energies
$\varepsilon^*_n$ and concomitant large peaks in the DOS.

\begin{figure}[hbt]
\includegraphics[scale=0.35]{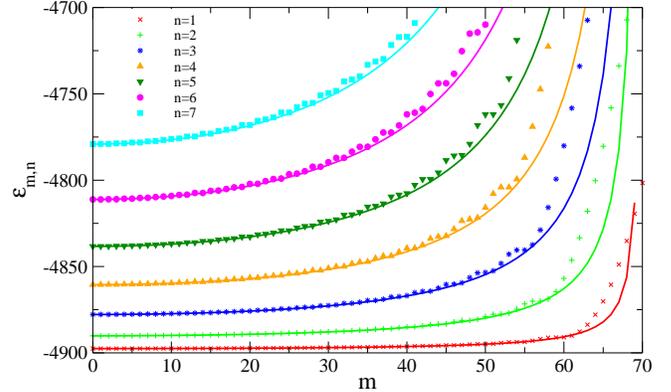}
\caption{(Color online) The $m$ dependence of the exact energy levels 
(in units of $\hbar^2/2m^*R^2$) of circular Rashba billiards (symbols) 
for a given $n$ ranging from $n=1$ to $n=7$. 
The solid lines are the curves obtained from the approximation of the
exact energy levels given by Eq.~(\ref{e_sing:eq}) 
as functions of $m$ with the corresponding $n$.  
Here $k_{\text{so}} R =70$. 
\label{E_m:figure}}
\end{figure}

To see the reason for this behavior,
it is useful to plot the energy levels as functions of $m$, as shown 
in Fig.~\ref{E_m:figure}.   
The curves in the figure start almost horizontally at $\varepsilon^*_n$, 
$n=1,2,\ldots$ resulting in large peaks in the DOS at the same energies.
Using Debye's asymptotic expression for Bessel functions with large
argument~\cite{Abramowitz-Stegun:konyv}, we were able to derive the
energy dispersion in leading order:
\begin{equation}
\varepsilon_{m,n} = \varepsilon_{\text{so}} \, 
\Biggl[
\frac{{(\frac{n\pi}{2})}^2}{\varepsilon_{\text{so}}- m^2} -1 
\Biggr]
 \label{e_sing:eq} 
\end{equation}
valid only for negative energies.
Figure~\ref{E_m:figure} also shows the comparison of the exact energy
levels and their approximated $m$ and $n$ dependence given 
by Eq.~(\ref{e_sing:eq}).
For small $m,n$ the above expression agrees excellently 
with the numerics (e.g., $\varepsilon_{0,1}$ is accurate up
to 7 digits for $\varepsilon_{\text{so}}=70$).
The smallest energy level in the spectrum of the circular Rashba billiard 
is $E_{\text{min}} = \hbar^2 /(2 m^* R^2) \, \varepsilon_{0,1}  
\cong \hbar^2/(2m^*)\, \pi^2/(4R^2) -\Delta_{\text{so}}$. 

We now derive an approximated expression for the counting function using  
\begin{equation}
N_{\text{asymp}}(\varepsilon)  = 2\, \sum_{m=0}^{m_{\text{max}}} 
\sum_{n=1}^{n_{\text{max}}} \, 
\Theta (\varepsilon- \varepsilon_{m,n}), 
\label{N_E-def:eq}
\end{equation}
where $\varepsilon_{m,n}$ are given by Eq.~(\ref{e_sing:eq}), 
the factor 2 takes into account the Kramers degeneracy in $m$, and 
$m_{\text{max}} = [\sqrt{\varepsilon_{\text{so}}}\,]$ and 
$n_{\text{max}}= [(2 \sqrt{\varepsilon_{\text{so}}}/\pi ]$ are 
the largest $m$ and $n$ for which $\varepsilon_{m,n}$ is still negative.  
Here $\left[\cdot \right]$ stands for the integer part. 
Applying the Poisson summation formula~\cite{Berry_Chaos:book,Brack:konyv} 
in the sum over $m$ in Eq.~(\ref{N_E-def:eq}) 
and keeping only the non-oscillating term we find 
\begin{eqnarray}
N_{\text{asymp}}(\varepsilon) &=& 2 \sum_{n=1}^{n_{\text{max}}} \, 
\int_{-\frac{1}{2}}^{m_{\text{max}}+\frac{1}{2}} \, 
\Theta (\varepsilon-\varepsilon_{m,n})\, dm 
\nonumber \\
&=& 2 \sum_{n=1}^{n_{\text{max}}} \, m^* (\varepsilon,n),
\label{N_asymp-1:eq}
\end{eqnarray}
where $m^*(\varepsilon,n)$  is the solution of 
$\varepsilon_{m,n} = \varepsilon$  for $m$ at a given $\varepsilon$ and $n$. 
Thus, from Eq.~(\ref{N_asymp-1:eq}), after some simple algebra, 
we obtain the final form of the asymptotic
counting function in Debye's approximation:
\begin{equation}
N_{\text{asymp}}(\varepsilon) = 2 \, \sqrt{\varepsilon_{\text{so}}}\, 
\sum_{n=0}^{n_{\text{max}}}\,  
\sqrt{\frac{\varepsilon -\varepsilon^*_{n}}
{\varepsilon +\varepsilon_{\text{so}}}} \, 
\Theta(\varepsilon -\varepsilon^*_{n}), 
\,\, \text{for} \,\, \varepsilon <0,
\label{N_asymp-2:eq}
\end{equation}
where $\varepsilon^*_{n} = \varepsilon_{0,n} 
= {(\frac{n\pi}{2})}^2-\varepsilon_{\text{so}}$. 
The result is plotted together with the exact counting function 
in Fig.~\ref{N_E-rounded:figure}. 
The agreement is clearly visible near the bottom of the spectrum.
However, it is an open question what semi-classical picture 
can be associated to the content of Eq.~(\ref{e_sing:eq}). 
A possible treatment in this direction may be the  
semi-classical approach of Ref.~\onlinecite{Pletyukhov-Brack:cikk}.  

The density of states is the derivative of the counting function
$N(E)$ with respect to $E$, therefore for circular Rashba billiards 
in the DOS square root types singularities (van Hove type) appear 
at energies $E^{\text{sing}}_n = \hbar^2 /(2 m^* R^2) \, \varepsilon^*_n $. 
Attaching leads to a circular Rashba billiard, the transport
properties of this open system are determined by 
the tunneling conductance which is proportional to the DOS.  
Thus, the measured conductance should be changed abruptly 
in the negative energy spectrum of circular Rashba billiards 
at energies $E^{\text{sing}}_n $. 

\subsection{The spin structures of the eigenstates}
\label{spin_circ:ch}

It is straightforward to obtain
corresponding spinor eigenstates and calculate their expectation value for the
$z$ component of spin. Similar to the case of Rashba--split eigenstates in
rings~\cite{Uli_persistent:cikk}, but in contrast to that of quantum
wires~\cite{hausler,Uli-1:cikk}, it turns out to be finite. 

The eigenstates of the Rashba billiards satisfying the Dirichlet
boundary conditions can be expressed with the linear combination of the 
regular eigenspinors $\mid \chi_m^{\pm} \rangle$ given by (\ref{eq:chi}):   
\begin{eqnarray}
\lefteqn{\Psi_{m,n}(r,\varphi) 
= \frac{1}{\sqrt{\cal{N}}} \, \Biggl \{ 
c_+ \left(\begin{array}{c}  J_m(k_+ r) \\ 
J_{m+1}(k_+ r)\, e^{i\varphi}
\end{array}
\right)  } \nonumber \\
& & +  c_- \left(\begin{array}{c} - J_m(k_- r) \\ 
J_{m+1}(k_- r)\,  e^{i\varphi}
\end{array}
\right)  \Biggr\} \, e^{im \varphi}  , 
\label{eigenspinor:eq}
\end{eqnarray}
where $\cal{N}$ is the normalization constant, the coefficients
$c_{\pm}$ 
satisfy 
\begin{equation}
\frac{c_+}{c_-} = \frac{J_m(k_-R)}{J_m(k_+R)} = 
- \frac{J_{m+1}(k_-R)}{J_{m+1}(k_+R)} ,
\label{a_per_b:def} 
\end{equation}
and $k_\pm$ satisfy the secular equation (\ref{sec_circ:eq})
with energy levels $\varepsilon_{m,n}$. 
Eigenstates given by Eq.~(\ref{eigenspinor:eq}) are valid for 
$\varepsilon_{m,n}> 0$. 
In the opposite case, one should use the eigenspinor given in
(\ref{eq:chi}) for $E <  0$.
Regarding the spin structures, it turns out that both cases 
(the positive and negative energy levels) can be treated at the same level 
if the definitions for $k_\pm$ in (\ref{kpm:def}) are modified as 
$k_{\pm} = k \mp k_{\text{so}}$.
Therefore, hereafter we use these new definitions for $k_\pm$.

The spin structure of the eigenstates in Rashba billiards 
can be obtained by calculating the expectation values 
for spin components: 
\begin{equation}
 \big\langle\sigma_i\big\rangle_{m,n} =
\int_0^R \int_0^{2\pi}  \!\!  r dr d\varphi 
\Psi_{m,n}^{+}(r,\varphi) \, \sigma_i \, \Psi_{m,n}(r,\varphi), 
\label{spin_struc-1:eq}
\end{equation}
where $i= x,y,z$, and $+$ denotes the transpose and the complex
conjugation of a spinor state.
The integrand in this equation is the spin density of $\sigma_i $.  
The eigenstates (\ref{eigenspinor:eq}) can be written in the form of  
\begin{equation}
\Psi_{m,n}(r,\varphi) = 
\left(\begin{array}{c} \Psi_{m,n}^{(1)}(r) \\[1ex] 
\Psi_{m,n}^{(2)}(r) e^{i \varphi} 
\end{array} \right)\, e^{i m\varphi}, 
\end{equation}
and then it is easy to show that the spin density depends only on $r$
as  
\begin{subequations} 
\begin{eqnarray}
\Psi_{m,n}^{+}({\bf r}) \, \sigma_z \, \Psi_{m,n}({\bf r}) 
 &=&
\!\! {\left|\Psi_{m,n}^{(1)}(r)\right|}^2  
\!  - \!  {\left|\Psi_{m,n}^{(2)}(r)\right|}^2 \!\! , \\[2ex]
\Psi_{m,n}^{+}({\bf r}) \, \sigma_r \, \Psi_{m,n}({\bf r}) 
&=&
2 \Psi_{m,n}^{(1)}(r) \Psi_{m,n}^{(2)}(r) , 
\label{spin_struc-2:eq}
\end{eqnarray}%
\end{subequations}% 
where $\sigma_r = \cos \, \varphi \sigma_x + \sin \, \varphi \sigma_y$
is the in-plane radial component of the spin.
One can also show that the angular component 
$\Psi_{m,n}^{+} \, \sigma_\varphi \, \Psi_{m,n}$ of 
the in-plane spin density is exactly zero, where 
$\sigma_\varphi = -\sin \varphi \, \sigma_x + \cos \varphi \,\sigma_y$. 
Therefore, the in-plane spin density at point ${\bf r}$ in the
billiard is along the radial direction
${\bf r}$~\cite{Rodriguez}.
This implies that the expectation values for in-plane spin is zero, i.e., 
$\langle\sigma_x \rangle_{m,n} = \langle\sigma_y \rangle_{m,n} = 0$. 

Performing the integration (that can be carried out analytically) 
in Eq.~(\ref{spin_struc-1:eq}) for $\langle\sigma_z \rangle_{m,n}$ 
we find
\begin{widetext}
\begin{eqnarray}
\big\langle\sigma_z\big\rangle_{m,n} &=& 
-\frac{\varepsilon_{m,n} 
+ \varepsilon_{\text{so}}}{\sqrt{\varepsilon_{\text{so}}}}
\, \frac{1}{\left[\frac{J_m(k_- R)}{J_{m+1}(k_- R)}+\frac{J_{m+1}(k_-
R)}{J_m(k_- R)}\right]\varepsilon_{m,n} 
+ \left(2m+1 \right)\, \sqrt{\varepsilon_{\text{so}}} }.
\label{sz-exact:eq}
\end{eqnarray}
\end{widetext}
This is an exact analytic result for the expectation values of the
$z$ component of the spin for circular Rashba billiards.

\begin{figure}[htb]
\includegraphics[scale=0.7]{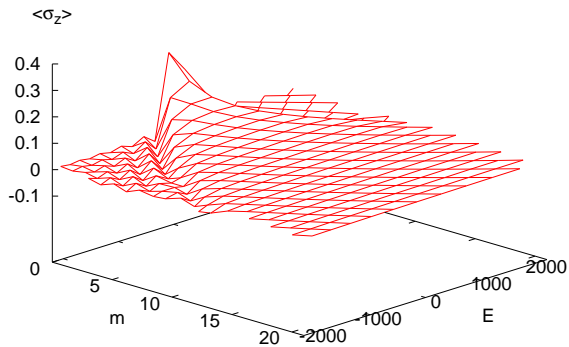}
\caption{The expectation values of $\big\langle\sigma_z\big\rangle_{m,n}$
as functions of the angular quantum number $m$ and 
eigenvalues $\varepsilon_{m,n}$ for $k_{\text{so}} R =70$.
\label{spinz-m-E:fig}}
\end{figure}

\begin{figure}[htb]
\includegraphics[scale=0.7]{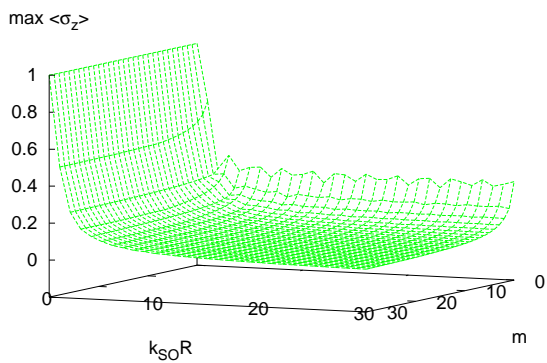}
\caption{The maximum of the 
$\big\langle\sigma_z\big\rangle_{m,n}$ as functions of $m$ and 
$X_{\text{so}}=k_{\text{so}} R$. 
\label{robustness-Sz:fig}}
\end{figure}

Figure~\ref{spinz-m-E:fig} shows the expectation values of 
$\big\langle\sigma_z\big\rangle_{m,n}$ 
calculated numerically from (\ref{sz-exact:eq}) 
for different angular quantum number $m$ and eigenvalues $\varepsilon_{m,n}$ 
with a given Rashba coupling strength $\alpha$. 
One can see from the figure that
$\big\langle\sigma_z\big\rangle_{m,n}$ has a peak at $m=0$ and for
eigenvalue $\varepsilon_{m,n}$ close to zero.
We have studied how this peak value changes for different Rashba
coupling strength $\alpha$.
For each $k_{\text{so}} R$ and $m$ the maximum of
$\big\langle\sigma_z\big\rangle_{m,n}$ over the eigenvalues 
$\varepsilon_{m,n}$  is plotted in Fig.~\ref{robustness-Sz:fig}. 
It is clear from the figure that the expectation values of
$\big\langle\sigma_z\big\rangle_{m,n}$ is robust for 
different Rashba coupling strength $\alpha$.

For Rashba billiards, in weak magnetic field 
the energy levels of the Kramers doublets 
will be splitted by the Zeeman effect. 
Using the first order perturbation valid for weak field limit, i.e.,
when the cyclotron radius is much larger than the size of the Rashba
billiard, the values of the Zeeman splitting is 
proportional to the expectation values of
$\big\langle\sigma_z\big\rangle_{m,n}$. 
Thus, we believe that the significant magnitude of the spin $z$
component found from our numerical results can be detectable experimentally.

\section{Conclusions}
\label{conclusion:ch}

Before concluding in this section we briefly summarize our results 
not discussed in this paper on the statistics of energy levels, 
and highlight some open theoretical problems in connection 
with Rashba billiards. 

The Schr{\"o}dinger equation (including boundary conditions) 
for circular Rashba billiards is separable
in polar coordinates, thus integrable. Hence, the statistics of energy levels
should be Poissonian (see e.g.~Ref.~\onlinecite{Q-billiard:konyv}).
Indeed, we have found that the nearest--neighbor 
level--spacing distribution $P(s)$ is Poissonian (not shown here).
For other shapes of Rashba billiards, spin-orbit coupling may destroy
integrability, in which case Random Matrix Theory (RMT) predicts that the
level statistics should be governed by the symplectic 
ensemble~\cite{Q-billiard:konyv,Carlo-RMT}. 
Note, however, that some intermediate distribution (not described by RMT) was
found~\cite{Berggren:cikk} for a rectangularly shaped billiard in the limit of
small $k_{\text{so}}$, reflecting the fact that a rectangular billiard is
integrable in the absence of SO coupling but non-integrable when SO is finite.

We now list a few interesting open theoretical problems. 
The Weyl formula is essential to develop a periodic orbit theory 
for Rashba billiards. (For normal billiards, see Brack and Bhaduri's book 
in Ref.~\onlinecite{Baltes-Hilf:book,Brack:konyv}, and a theory in
case of harmonically confined Rashba systems is given in
Ref.~\onlinecite{Pletyukhov-Brack:cikk}.) The Green's function method
presented in this work would be a suitable starting point to calculate
observables such as the magnetization~\cite{Silvia-Rocca-1:cikk} or 
persistent currents~\cite{Uli_persistent:cikk} in Rashba billiards. 

In conclusion, we have presented a study of electron billiards with
spin--dependent dynamics due to Rashba spin splitting.
Semi-classical results for the spectrum agree well with exact
quantum calculations. We find interesting properties of negative--energy
states, including a finite spin projection in the out--of--plane
direction.

\acknowledgments

This work is supported in part by E.~C.\ Contract No.~MRTN-CT-2003-504574, 
and the Hungarian  Science Foundation OTKA T034832, T046129 and T038202. 
U.~Z. gratefully acknowledges funding from the Marsden Fund of the Royal
Society of New Zealand.


\begin{thebibliography}{xx}

\bibitem{spintronics-book}
{\it Semiconductor Spintronics and Quantum Computation}, 
edited by D. D. Awschalom, D. Loss, and N. Samarth (Springer, Berlin, 2002). 

\bibitem{Wolf-review:cikk}
S.~A.~Wolf, D.~D.~Awschalom, R.~A.~Buhrman, J.~M.~Daughton, 
S.~von~Moln\'ar, M.~L.~Roukes, A.~Y.~Chtchelkanova and D.~M.~Treger, 
Science  {\bf 294}, 1488 (2001).

\bibitem{roessler}
G. Lommer, F. Malcher, and U. R\"ossler, Phys. Rev. Lett.
{\bf 60}, 728 (1988).

\bibitem{Rashba:cikk} 
E. I. Rashba, Fiz. Tverd. Tela (Leningrad)  {\bf 2}, 1224 (1960)
  [Sov. Phys. Solid State {\bf 2}, 1109 (1960)].

\bibitem{nitta}
J.~Nitta, T.~Akazaki, H.~Takayanagi, and T.~Enoki, Phys. Rev.
Lett. {\bf 78}, 1335 (1997). 

\bibitem{thomas}
T.~Sch\"apers, G.~Engels, J.~Lange, Th.~Klocke, M.~Hollfelder, and H.~L\"uth, 
J. Appl. Phys. {\bf 83}, 4324 (1998).  

%UZ: Added reference below
\bibitem{syoji:jap:01}
Y. Sato, T. Kita, S. Gozu, and S. Yamada, J. Appl. Phys. {\bf 89}, 8017 (2001).

\bibitem{Datta-Das:cikk} 
S. Datta and B. Das, Appl. Phys. Lett.  {\bf 56}, 665 (1990);

\bibitem{Mireles-Kirczenow:cikk}
F. Mireles and G. Kirczenow, Phys. Rev. B  {\bf 64}, 024426 (2001).

\bibitem{hausler}
W. H\"ausler, Phys. Rev. B {\bf 63}, 121310 (2001).

\bibitem{Uli-1:cikk}
M.  Governale and U. Z\"ulicke, Phys. Rev. B  {\bf 66}, 073311 (2002).

\bibitem{thomas_wire}
T. Sch\"apers, J. Knobbe, and V.~A. Guzenko, Phys. Rev. B {\bf 69}, 235323
(2004).

\bibitem{Uli_persistent:cikk}
J. Splettstoesser, M. Governale, and U. Z\"ulicke, 
 Phys. Rev. B {\bf 68}, 165341 (2003).  

\bibitem{Benedict-1:cikk}
P. F\"oldi, B. Moln\'ar, M. G. Benedict, and F.~M.~Peteers, 
Phys. Rev. B {\bf 71}, 033309 (2005).

\bibitem{dot1}
O. Voskoboynikov, C.~P. Lee, and O. Tretyak, Phys. Rev. B {\bf 63}, 165306
(2001).

\bibitem{Michele-q-dot:cikk}
M. Governale, Phys. Rev. Lett.  {\bf 89}, 206802 (2002).

\bibitem{dot2}
M. Val\'{i}n-Rodr\'{i}guez, A. Puente, and L. Serra, Phys. Rev. B {\bf 69},
085306 (2004).

\bibitem{Bulgakov-Sadreev:cikk}
E. N. Bulgakov and A. F. Sadreev, Phys. Rev. B {\bf 66}, 075331 (2002). 

\bibitem{Rodriguez} M. Val\'{i}n-Rodr\'{i}guez, A. Puente, and L. Serra, 
Phys. Rev. B,  {\bf 69}, 153308 (2004). 

\bibitem{dot3}
C.~F. Destefani, S.~E. Ulloa, and G.~E. Marques, Phys. Rev. B {\bf 69},
125302 (2004).

\bibitem{zaitsev}
O. Zaitsev, D. Frustaglia, and K. Richter, 
Phys. Rev. Lett.  {\bf 94}, 026809 (2005).

%UZ: Added 2004 Sadreev reference below
\bibitem{Sadreev:cikk} E. N. Bulgakov and A. F. Sadreev, JETP Letters,
{\bf 78}, 443 (2003); Phys. Rev. E {\bf 70}, 056211 (2004); A. I. Saichev,
H. Ishio, A. F. Sadreev and K.-F. Berggren, J. Phys. A {\bf 35}, L87 (2002). 

\bibitem{Berggren:cikk} K.-F. Berggren an T. Ouchterlony, Found. Phys.
{\bf 31}, 233 (2001).

%UZ: Added reference below.
\bibitem{falko}
I.~L. Aleiner and V.~I. Fal'ko, Phys. Rev. Lett. {\bf 87}, 256801 (2001);
J.-H. Cremers, P.~W. Brouwer, and V.~I. Fal'ko, Phys. Rev. B {\bf 68}, 125329
(2003).

\bibitem{bychkov}
Yu.~A. Bychkov and E.~I. Rashba, J. Phys. C {\bf 17}, 6039 (1884).

\bibitem{Weyl:cikk} 
H. Weyl, G{\"o}ttinger Nachrichten  {\bf 110}, 114 (1911).

\bibitem{Baltes-Hilf:book}  
H. T. Baltes and E. R. Hilf, {\it Spectra of Finite Systems}
(Bibliographisches Institut Wissenschaftsverlag, Mannheim, 1976); 

\bibitem{Brack:konyv} M. Brack and R. K. Bhaduri, {\it Semiclassical Physics} 
(Addison-Wesley, Reading, 1997).

\bibitem{Kac:cikk} 
M. Kac, Am. Math. Monthly  {\bf 73}, 1 (1966).

\bibitem{Balian-Bloch:cikkek} 
R. Balian and C. Bloch, Ann. Phys. (N.Y.)  {\bf 60}, 401 (1970); 

\bibitem{Stewartson-Waechter:cikk} 
K. Stewartson and R. T. Waechter, 
Proc. Cambridge Philos.\ Soc.\  {\bf 69}, 581 (1971).

\bibitem{Berry-Howls:cikk} 
M. Berry and C.J. Howls, 
Proc.\ R.\ Soc.\ Lond.\ A {\bf 447}, 527 (1994). 

\bibitem{Uzy-Sieber:cikk} 
M. Sieber, H. Primack, U. Smilansky, I. Ussishkin, and H. Schanz, 
J. Phys. A  {\bf 28}, 5041 (1995). 

\bibitem{Q-billiard:konyv}
{\it Chaos and Quantum Physics}, ed.~by 
M.-J. Giannoni, A. Voros and J. Zinn-Justin 
(Elsevier Science Publishers B.V., Amsterdam, The Netherlands, 1991).
(Elsevier, Amsterdam, The Netherlands, 1991).

\bibitem{Berry-neutrino:cikk} 
M. V. Berry and R. J. Mondragon, 
Proc.\ R.\ Soc.\ Lond.\ A {\bf 412}, 53 (1987).

\bibitem{Bulgakov-seq_eq}
E. N. Bulgakov and A. F. Sadreev,  JETP Letters, {\bf 73}, 505 (2001).

\bibitem{magneses-dot_SO-exact:cikk} 
E. Tsitsishvili, G. S. Lozano and A. O. Gogolin, 
Phys. Rev. B {\bf 70}, 115316 (2004).  

\bibitem{malshukov}
C.-H. Chang, A.~G. Mal'shukov, and K.~A. Chao, 
Phys. Rev. B {\bf 70}, 245309 (2004).

\bibitem{JAU:paper} J. Cserti, A. Csord\'as and U. Z\"ulicke,  
Phys. Rev. B {\bf 70}, 233307 (2004).

\bibitem{Pletyukhov-Brack:cikk} 
M. Pletyukhov, Ch. Amann, M. Mehta, and M. Brack, 
Phys. Rev. Lett. {\bf 89}, 116601 (2002); 
Ch. Amann and M. Brack, J. Phys. A {\bf 35}, 6009 (2002).

\bibitem{Walls_Heller:cikk} 
J. D. Walls, J. Huang, R. M. Westervelt, E. J. Heller, cond-mat/0507528.

\bibitem{Abramowitz-Stegun:konyv}  
M. Abramowitz and I. A. Stegun, {\it Handbook of Mathematical Functions} 
(Dover, New-York, 1972).

\bibitem{Gradshteyn-Ryzhik:konyv} 
I. S. Gradshteyn and I. M. Ryzhik, {\it Table of Integrals, Series, and 
Products}, 5th ed. (Academic Press, San Diego, 1994).

\bibitem{Berry_Chaos:book} M. Berry in
  Ref.~\onlinecite{Q-billiard:konyv}, p.~251. 

\bibitem{ray-splitting:cikk} 
Y. D{\'e}canini and A. Folacci, Phys. Rev. E  {\bf 68}, 046204 (2003).

\bibitem{Schmit:cikk_Csordas:cikk}
C. Schmit in Ref.~\onlinecite{Q-billiard:konyv}, p.~331;
A.~Csord\'as, R.~Graham, P.~Sz{\'e}pfalusy,
Phys.\ Rev.\ A {\bf 44}, 1491 (1991).

\bibitem{Carlo-RMT} 
C. W. J. Beenakker, Rev. Mod. Phys. {\bf 69}, 731 (1997).

\bibitem{Silvia-Rocca-1:cikk} E. A. de Andrada e Silva, G. C. La Rocca, and
F. Bassani, Phys. Rev. B {\bf 50}, 8523 (1994).   

\end{thebibliography}
\end{document}